\documentclass[journal]{IEEEtran}
\usepackage[pdftex]{graphicx}

\graphicspath{{./pdf/}}

\DeclareGraphicsExtensions{.pdf}

\usepackage[cmex10]{amsmath}
\usepackage{cite}

\usepackage{url}

\begin{document}
\title{Absolute light yield measurements on SrF$_{2}$ and BaF$_{2}$ doped with rare earth ions}

\author{Roman~Shendrik,~\IEEEmembership{Member,~IEEE,} and
        Evgeny~Radzhabov

\thanks{Roman Shendrik and Evgeny Radzhabov are with the Laboratory
of Physics of monocrystals, Vinogradov Institute of Geochemistry SB RAS, Irkutsk,
664033, Favorskogo street, 1a, Russia, and Irkutsk State University, blvd Gagarina 20, Irkutsk, Russia, 664003, e-mail: shendrik@ieee.org}% $<$-this % stops a space
}

\markboth{Journal of \LaTeX\ Class Files,~Vol.~6, No.~1, January~2012}%
{Shell \MakeLowercase{\textit{et al.}}: Bare Demo of IEEEtran.cls for Journals}

\maketitle

\begin{abstract}

Results of absolute light output measurements on strontium and barium fluoride doped with PrF$_3$ and CeF$_3$ are presented and compared with scintillators having well-known light output (NaI-Tl, CsI-Tl, BGO). For pure SrF$_2$ crystal we obtain a value of about 28600 photons/MeV.
\end{abstract}


\IEEEpeerreviewmaketitle

\section{Introduction}

\IEEEPARstart{I}n previous investigations, we have found that SrF$_2$ and BaF$_2$ crystals doped with Ce$^{3+}$ and Pr$^{3+}$ ions are prospective scintillators for well-logging applications \cite{Shendrik2010, Shendrik2012, Shendrik2013}. Next step in scintillation properties study of these crystals is absolute light output measurements. 

The measurement of the absolute light yield of a scintillator is complicated. Not all photons created by ionizing radiation in the scintillation crystal are reached a photocathode of photomultiplier (PMT) that reduces number of photoelectrons in PMT in which photons are converted. Photon losses can be caused by a variety of reasons: the size of crystal, surface quality of scintillator, sensitivity of photocathode in scintillator emission region, and finally photoelectron collection efficiency. 

In some papers methods of absolute light output measurements were presented \cite{Holl88, Dorenbos93, Moszynski1997, Haas05, Gierlik07, Haas08, Moszynski10}. Light output of popular scintillators, e.g. NaI-Tl, CsI-Tl, BGO, is measured in the most papers. Investigation of fluoride scintillators is not in the spread. First spread fluoride scintillator was CaF$_{2}$-Eu. Its absolute light output was evaluated as 24000 photons/MeV \cite{Holl88}. The next explosion of interest in fluoride scintillators was in barium fluoride crystal. It is relatively dense material with shortest decay time (about 0.8 ns).
Absolute light yield was measured on BaF$_2$ and BaF$_2$-Ce crystals \cite{Melcher89, Visser1991, Dorenbos93, Janus09}. Typical light yield of pure BaF$_2$ (decay constant 600 ns) was about 9500-10000 photons/MeV \cite{Dorenbos93, Janus09}. Fast component (0.8 ns) light output was about 1500 photons/MeV \cite{Melcher89}. Also light output of BaF$_{2}$ doped with Ce$^{3+}$ was estimated about 13000 photons/MeV \cite{Visser1991}. Absolute light yield of SrF$_2$, SrF$_2$-Ce, and SrF$_2$-Pr crystals has not been measured yet. Nevertheless, preliminary evaluations give light yield of pure SrF$_2$ about 30000 photons/MeV \cite{Shendrik2013}. 

The aim of this paper was to study the light output of different scintillators using calibrated Enterprises 9814QSB photomultiplier. Photoelectron yield is determined from comparison full-energy peak of scintillator with single electron response. The measured numbers of photoelectrons were converted into photon number using the quantum efficiency of the PMT as specified by the manufacturer. 

Measurements were made for samples of popular scintillators BGO, CsI-Tl, and NaI-Tl and some pure and Pr-doped or Ce-doped fluoride scintillators. There are SrF$_2$ and BaF$_2$ crystals doped with Pr$^{3+}$ and Ce$^{3+}$ ions in concentrations of 0.01, 0.1, 0.3, and 1 mol.\%. The given concentrations of dopants were chosen because crystals doped with higher than 1 mol.\% concentrations of Pr$^{3+}$/Ce$^{3+}$ ions showed dramatically low light outputs \cite{Radzhabov2012Arx, Gektin2009}. 
\begin{table*}[!t]
\renewcommand{\arraystretch}{1.3}
\begin{minipage}{\textwidth}
\centering
\caption{Tested crystals}
\label{table-dim}
\begin{tabular}{llllll}
\hline
\hline
Sample & $\rho$ & Refractive index & Cleavage plane & Size & Manufacturer \\
 & [g/cm$^3$] & @ emission max\footnote{Data were given by \cite{RefInd}} & & [mm] & \\

\hline

NaI-Tl & 3.67 & 1.83 & $<$100$>$ & cylindrical, \O40x40 & "Crystal", Usol'e-Sibirskoe, Russia\\

CsI-Tl & 4.51 & 1.79 & none  & cubic, 10x10 & "Crystal", Usol'e-Sibirskoe, Russia \\

BGO (Bi$_{4}$Ge$_{3}$O$_{12}$) & 7.13 & 2.13 & none  & cylindrical, \O25x25 & NIIC SB RAS, Novosibirsk, Russia \\

BaF$_2$ & 4.88 & 1.50 & $<$111$>$ & cubic, 25x25 & IGC SB RAS, Irkutsk, Russia\\

BaF$_2$-Pr$^{3+}$ & 4.88 & 1.51 & $<$111$>$ & cylindrical, \O10x2 & IGC SB RAS\\
 & & & & cylindrical, \O25x25 & \\

SrF$_2$ & 4.18 & 1.46 & $<$111$>$ & cubic, 10x10 & IGC SB RAS\\

SrF$_2$ \footnote{Cleaved sample} & 4.18 & 1.46 & $<$111$>$ &  cylindrical, \O10x10 & IGC SB RAS\\

SrF$_2$-Ce$^{3+}$ & 4.18 & 1.46 & $<$111$>$ &  cylindrical, \O10x10 & IGC SB RAS\\

SrF$_2$-Pr$^{3+}$ & 4.18 & 1.50 & $<$111$>$ & cylindrical, \O10x2 &  IGC SB RAS\\

\hline
\hline
\end{tabular}
\end{minipage}
\end{table*}

\begin{table*}[!t]
\renewcommand{\arraystretch}{1.3}
\begin{minipage}{\textwidth}
\centering
\caption{Compilation of integral quantum efficiencies, wavelength of luminescence maxima, photoelectron ($Y_{phe}$) and absolute photon ($Y_{ph}$) yields, and full width at half maximum (FWHM) of tested scintillators in comparison with literature photoelectron yield data}
\label{table1}
\begin{tabular}{llllllll}
\hline
\hline
Scintillator & Primary & Wavelength of & Integral & Photoelectron & Reference & Photon & Full width at \\
 & decay time & emission max & quantum efficiency & yield & data $Y_{phe}$ & yield  & half maximum \\
 &  &  & QE$_{eff}$ & $Y_{phe}$ & & $Y_{ph}$ & FWHM  \\
\cline{2-8}
 &  [ns] &  [nm] & & [phe/MeV] & [phe/MeV] &  [ph/MeV] & [\%] \\

\hline

NaI-Tl & 250 & 415 & 0.21 & 8500$\pm$600 & 8900 \cite{Moszynski2006-1}& 38000 & 8 \\

CsI-Tl & 1000 & 540 & 0.07 & 4900$\pm$390 & 4400 \cite{Moszynski1997} & 55700 & 7.1 \\

BGO & 300  & 480 & 0.13 & 1380$\pm$100 & 1200 \cite{Moszynski1997} & 8200 & 15\\

BaF$_2$ & 600 & 280 & 0.21 & 1930$\pm$120 & 2110 \cite{Dorenbos93} & 9400 & 13 \\

BaF$_2$-0.15 mol.\% Pr$^{3+}$ & 21 & 228; 257 & 0.20 & 1230$\pm$80 & & 6300 & 23 \\

BaF$_2$-0.15 mol.\% Pr$^{3+}$ & 21 & 228; 257 & 0.20 & 1500$\pm$100 & & 7700 & 19 \\

SrF$_2$ & 1000 & 285 & 0.21 & 6010$\pm$420 & & 29200 & 10 \\

SrF$_2$ \footnote{Cleaved sample} & 1000 & 285 & 0.21 & 4020$\pm$280 & & 19500 & 13 \\

SrF$_2$-0.3 mol.\% Ce$^{3+}$ & 130\footnote{The crystal has long decays \cite{Shendrik2013}} & 310; 325 & 0.23 & 2100$\pm$150 & & 9300 & 13 \\

SrF$_2$-0.15 mol.\% Pr$^{3+}$ & 25 & 232; 250 & 0.19 & 2200$\pm$150 & & 11800 & \\

SrF$_2$-0.3 mol.\% Pr$^{3+}$ & 25 & 232; 250 & 0.19 & 1300$\pm$90 & & 7000 & 20\\

SrF$_2$-1 mol.\% Pr$^{3+}$ & 24 & 232; 250 & 0.19 & 220$\pm$20 & & 1200 & \\

\hline
\hline
\end{tabular}
\end{minipage}
\end{table*}

\section{Experimental methodology}

First, well-known scintillation crystals CsI-Tl, NaI-Tl, and BGO were tested. Crystals of CsI-Tl (10x10x10~mm$^3$) and NaI-Tl (40x40x40~mm$^3$) were given by Crystal, Usol'e, Russia. BGO (Bi$_4$Ge$_3$O$_{12}$) crystal was given by Yan Vasil'ev (Institute of Inorganic Chemistry, Novosibirsk, Russia). 

In second part of the study, fluoride crystals were tested. The crystals were grown using the Stockbarger method in graphite crucible in vacuum with 1 \% of CdF$_{2}$ added as a scavenger for oxygen containing impurities \cite{Nepomnyashchikh2001}. No fluoride crystals were contain oxygen that was controlled by absorption spectra. The samples were coated with PTFE tape to maximize the light collection efficiency and optically coupled to the window of PMT using glycerin grease. 

\begin{figure}[]
\centering
\includegraphics[width=3in]{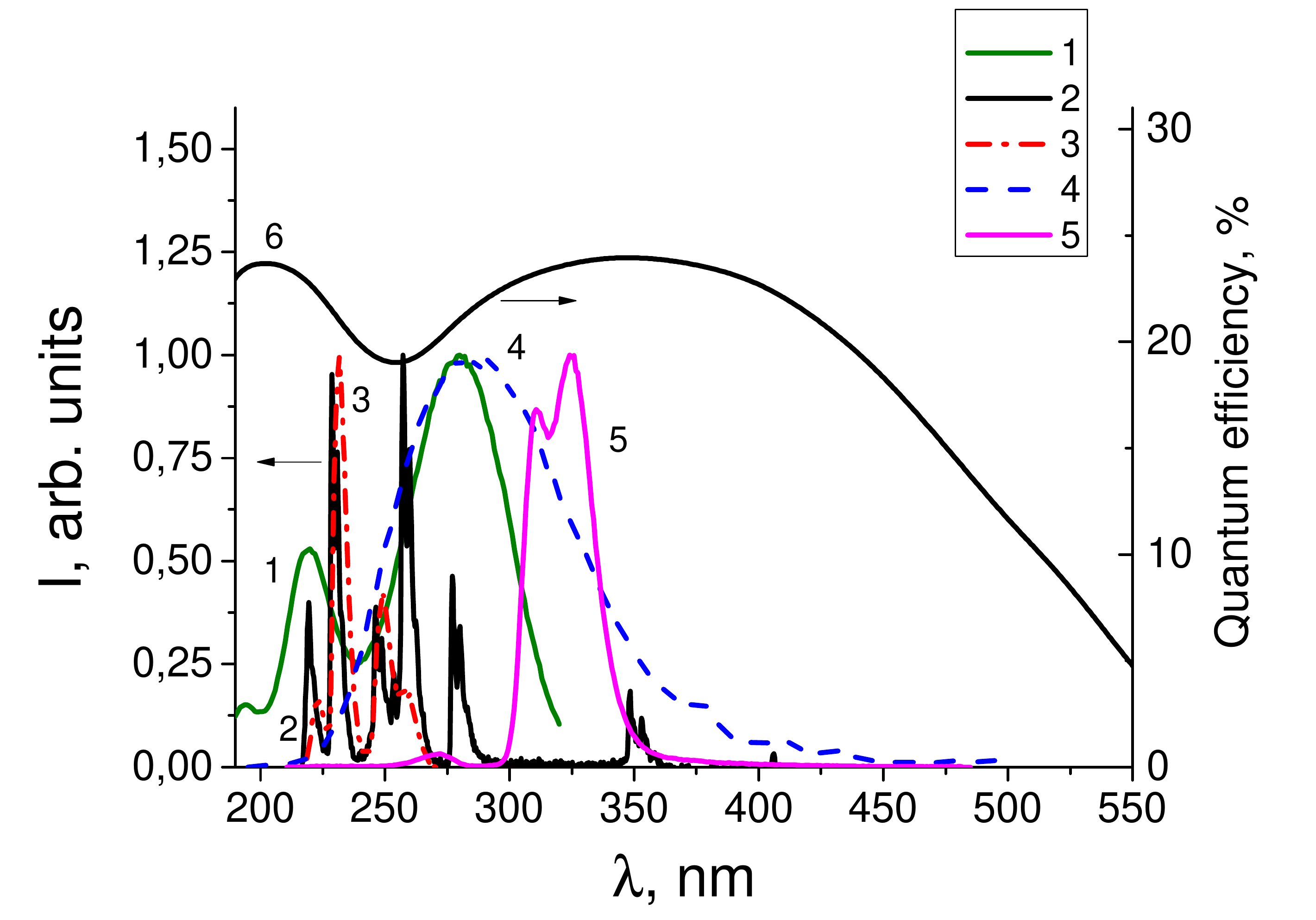}
\caption{Normalized emission spectra of pure BaF$_2$ (curve 1), BaF$_2$-Pr$^{3+}$ (curve 2), SrF$_2$-Pr$^{3+}$ (curve 3), pure SrF$_2$ (curve 4), and SrF$_2$-Ce$^{3+}$ (curve 5) in comparison with quantum efficiency of PMT 9814QSB (curve 6).}
\label{spectra}
\end{figure}

Luminescence of investigated fluoride crystals lies in UV spectral region. Therefore, photomultiplier with quartz windows is required. The most measurements of absolute light yield of the fluorides were carried out with PMT Photonis 2020Q. The photomultiplier tube Enterprises 9814QSB with comparable characteristics was used in our measurements. The quantum efficiency of photocathode were calibrated by manufacturer and given in Fig. \ref{spectra}, curve 6. This photomultiplier demonstrates smaller decrease of the sensitivity in 250-300 nm spectral region than PMT 2020Q. The PMT was operated with a CSN638C1 negative polarity voltage chain \cite{Enterprises}. The focusing system of 46 mm active diameter was assumed 100 \% photoelectron collection efficiency in the center of the photocathode. Diameter of maximal collection efficiency zone of photocathode of XP2020Q was estimated about 9 mm in paper \cite{Moszynski1997}. Measurements of photoelectron number were made three times to calculate the errors to the measured numerical quantities.

The crystal was irradiated with gamma rays from a monoenergetic $\gamma$-ray source of $^{137}$Cs  (E=662 KeV). A homemade preamplifier and an Ortec 570 amplifier were used to obtain the single electron and crystal pulse height spectra. The $^{137}$Cs pulse height spectrum of samples was determined for shaping times of 10 $\mu$s. The coarse gain control was used to switch between the single photoelectron pulse height spectrum and the $^{137}$Cs pulse height spectrum. The gain scale on the amplifier was linear.

X-ray excited luminescence was performed using X-ray tube with Pd anode operating at 35 kV and 0.8 mA. The spectra were recorded at photon-counting regime using PMT FEU-39A and vacuum grating monochromator VM-4.

\section{Results and discussion}
Luminescence spectra of investigated crystals are given in Fig.~\ref{spectra}. In the spectrum of SrF$_{2}$ a wide band at 280 nm are attributed to self-trapped exciton (STE) emission (Fig. \ref{spectra}, curve~4) \cite{Hayes_mon74}. In the luminescence spectrum of pure BaF$_{2}$ wide band corresponded to STE at 270-280 nm and a band peaked at 220 nm related to so-called core-valence transitions \cite{Rodnyi1991} or crossluminescence (Auger-free luminescence) are found (Fig. \ref{spectra}, curve~1).  

In Ce$^{3+}$-doped fluorides STE luminescence is quenched. The most intense bands in x-ray luminescence spectra of SrF$_{2}$-Ce$^{3+}$ crystals at 310 and 325 nm correspond to 5d-4f emission of Ce$^{3+}$ ions (fig. \ref{spectra}, curve~5). In Pr-doped crystals several bands observed in 220-360 nm spectral region are attributed to 5d-4f transitions in Pr$^{3+}$ ions \cite{Shendrik10}. 

In considering absolute light output integral quantum efficiency is important characteristic that has to be calculated taking account emission spectrum of a scintillator. Integral quantum efficiency values are given in Table~\ref{table1}. Almost all of the values of quantum efficiency lie in 0.19-0.23 interval. Emission of CsI-Tl and BGO crystals is observed in green spectral region, where sensitivity of the PMT is dropped. Therefore, integral quantum efficiencies for BGO and CsI-Tl crystals are 0.13 and 0.07, respectively.  

\begin{figure}[]
\centering
\includegraphics[width=3.8in]{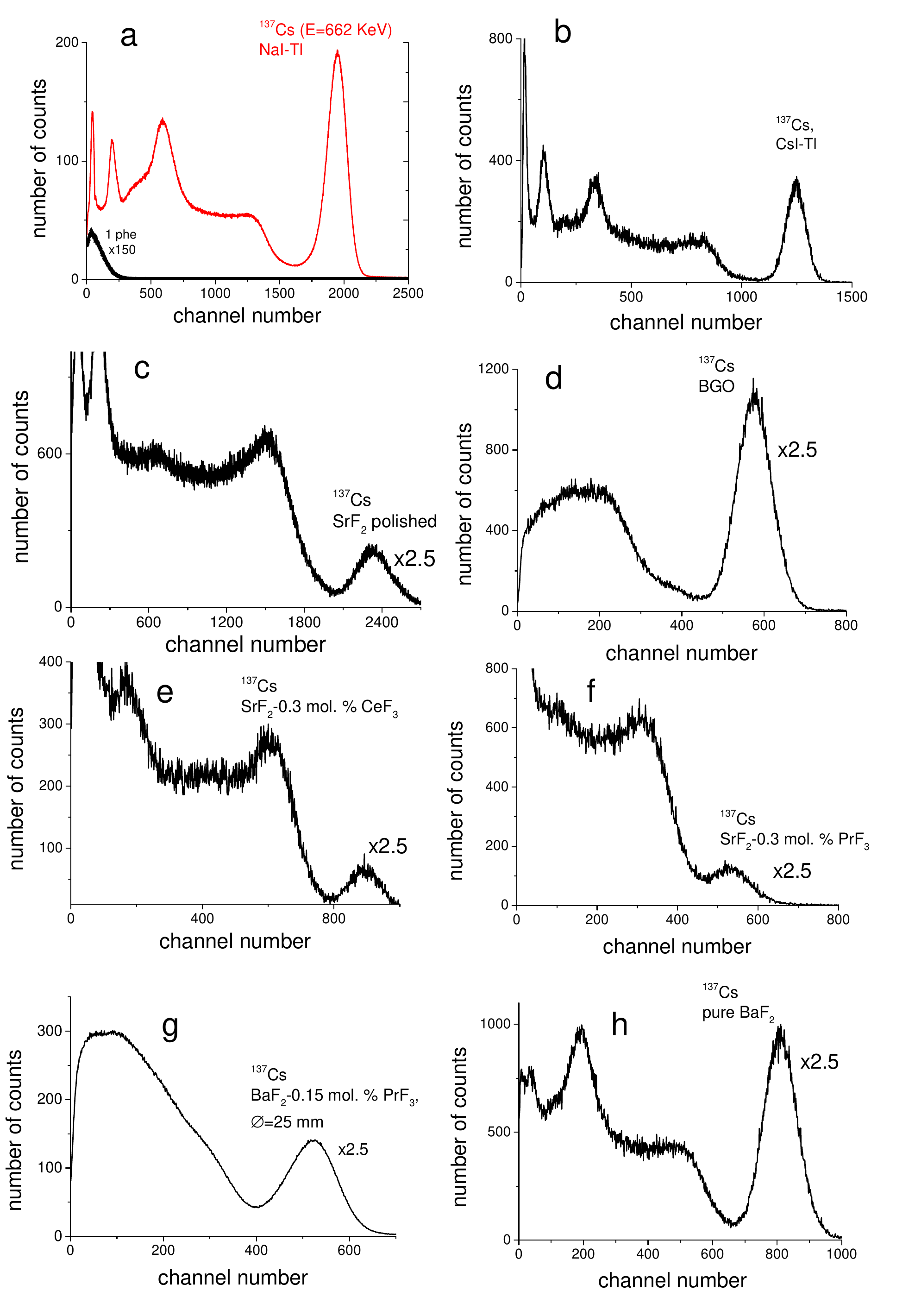}
\caption{A comparison of the energy spectra of $\gamma$-rays from a $^{137}$Cs source measured with different scintillators (a - NaI-Tl and single electron peak; b - CsI-Tl; c - pure SrF$_2$; d - BGO; e - SrF$_2$-0.3 mol.\% Ce$^{3+}$; f - SrF$_2$-0.3 mol.\% Pr$^{3+}$; g - BaF$_2$-0.15 mol.\% Pr$^{3+}$; h - pure BaF$_2$.).}
\label{pulse}
\end{figure}

Crystals CsI-Tl, NaI-Tl and BGO were used as testing scintillators with well-known photon and photoelectron light outputs ($Y_{phe}$). Fig.~\ref{pulse} shows pulse height spectra of pure and rare-earth doped fluorides, NaI-Tl, CsI-Tl, and BGO crystals. The photopeak corresponding to the $^{137}$Cs energy photon is seen in each curve in the Figure~\ref{pulse}. From the 662 keV photopeak in the pulse height spectrum and from the average pulse height in the single electron pulse height spectrum of the PMT, the photoelectron yield ($Y_{phe}$) is calculated Table \ref{table1}, column 5. NaI-Tl demonstrates highest $Y_{phe}$ about 8500 phe/MeV that agrees well with photoelectron output (8900 phe/MeV) measured by M. Moszynski et al \cite{Moszynski2006-1}. $Y_{phe}$ on CsI-Tl is about 4900 phe/MeV that is close to the 4400 reported by \cite{Moszynski1997}. Photoelectron yield of BGO -- 1380 phe/MeV is comparable with the yield measured in \cite{Moszynski1997}.
Photoelectron output of pure barium fluoride is 1930 phe/MeV that is lower than value 2110 phe/MeV given by P. Dorenbos et al \cite{Dorenbos93}. It should be noted, that in the cited papers measurements of photoelectron outputs were provided with PMT Photonis 2020Q. 

Photoelectron light output of polished pure SrF$_2$ crystal is 6010 phe/MeV. The cleaved crystal shows about 30\% lower photoelectron output than the polished crystal (4020 phe/MeV) due to lower light collection efficiency. After doping Ce$^{3+}$ or Pr$^{3+}$ ions photoelectron output of SrF$_{2}$ crystals is decreased. SrF$_{2}$ crystal doped with 0.15 mol.\% of Pr$^{3+}$ demonstrates the largest photoelectron output about 2200 phe/MeV. The output is decreased with increasing concentration of Pr$^{3+}$ ions. Photoelectron yield of SrF$_{2}$-1 mol.\% Pr$^{3+}$ is about 220 phe/MeV. Crystals doped with higher concentrations of Pr$^{3+}$ ions demonstrate low photoelectron output of several tens of photoelectrons/MeV.

Photoelectron output of BaF$_{2}$-Pr$^{3+}$ samples is lower than SrF$_{2}$-Pr$^{3+}$. The largest output (1500 phe/MeV) is observed for thin crystal of BaF$_{2}$-0.15 mol.\% Pr$^{3+}$ ($\O$10x2 mm). Photoelectron yield is decreased in the large size crystal ($\O$25x25 mm) due to less effective light collection and low quality of crystal surface.
 
To calculate the photon yield per MeV from the photoelectron yield/MeV, a correction must be made for the transmittance of the optical coupling compound, reflectivity losses, and the quantum efficiency of the employed PMT. To obtain the absolute photon yield $Y_{ph}$ we use the equation \cite{Bizarri2006}:
\begin{equation}
\label{LY}
Y_{ph}=Y_{phe}\frac{1-R_{eff}}{0.98*QE_{eff}},
\end{equation} 
where $QE_{eff}$ is the integral quantum efficiency given in Table \ref{table1}, $R_{eff}$ is the PMT effective reflectivity, and $Y_{phe}$ -- photoelectron yield. The light output of the
scintillator should be corrected for the reflectivity of the photocathode. In quantum efficiency measurement procedure a part of the light which is reflected from the photocathode is lost. In the
measurement with the crystal having full optical contact due to grease the reflected light can be collected back into the photocathode. Therefore, the effective quantum efficiency is increased. The largest effective reflectivity for PMT 9814QSB is observed in green spectral range. Such, the reflectivity in CsI-Tl luminescence region is about 21~\%, and it is reached 24~\% in BGO emission region. For NaI-Tl the reflectivity is given 8 \%. In UV spectral region no light is reflected from photocathode and $R_{eff}$ is less~1 \%. The parameters of effective reflectivity were measured by the PMT manufacturer.

Measured absolute light outputs of the tested samples (NaI-Tl, CsI-Tl, BaF$_{2}$ and BGO) are similar to known results \cite{Dorenbos93, Moszynski1997, Berkley}. The largest light output 29200 photons/MeV in fluoride crystals was found for pure SrF$_2$ crystals (Table~\ref{table1}, column 7). The results confirm the value given in our earlier publication \cite{Shendrik2013}. 

Fluoride crystals doped with trivalent rare-earth ions show light output lower than pure crystals due to ineffective energy transfer mechanism. The most favorable energy transfer in fluoride crystals is resonance STE transfer \cite{Visser1991, Radzhabov2012Arx}. This mechanism takes place at room temperature only in the crystals doped with Ce$^{3+}$ ions \cite{Shendrik10}. Disadvantage of excitonic transfer is slower than in electron-hole transfer emission decay. However, decay time constant becomes shorter with increasing of Ce$^{3+}$ ions concentration due to extension of probability of resonance STE -- Ce-ion energy transfer. 

Decay time of the Ce-luminescence equals 130 ns in SrF$_{2}$-0.3 mol.\% Ce$^{3+}$ and it becomes longer with decrease of Ce concentration \cite{Shendrik2013}. Decay time of Ce$^{3+}$ ions luminescence under optical excitation is about 30 ns, that is comparable with radiative 5d-4f recombination time in Ce$^{3+}$ ion \cite{Visser1991, Visser93, Radzhabov2004}. Under vacuum ultraviolet excitation at exciton and higher energies regions the decay of Ce-doped ﬂuorides became nonexponential \cite{Visser1991, Wojtowicz00-2}. Observed decay constant (130 ns) can be ascribed to resonance energy transition in nearest pairs of exciton and cerium ion \cite{Visser93, Radzhabov2012Arx}.
Electron-hole energy transfer is also found in Ce-doped fluoride crystals. However, delayed energy transfer, when electron or hole is captured by trap instead direct recombination in Ce-ions, is dominant \cite{Shendrik2013}. Notwithstanding the fact that in Ce-doped crystals eventual light output could be reached 35000 photons/MeV a large part of emitted light is not registered in pulse height measurements due to presence long time components (about tens and hundreds of microseconds) in Ce ions emission \cite{Shendrik2013}.

In Pr$^{3+}$ doped crystals consecutive capture of electron forming Pr$^{2+}$ center and then hole with following recombination is primary luminescence mechanism. Apart fast electron-hole recombination delayed energy transfer process involving hole traps takes place.  At room temperatures two compete processes are found. There are "prompt" electron-hole capture, when the activator ion traps consequently electron and hole, and "delayed" electron-hole capture, when the activator ion catches the electron and the hole coming to the activator via hole traps (V$_{kA}$). Herewith, the efficiency of the second process is a higher in the fluoride crystals  \cite{Shendrik2012}. Therefore, light output of Pr-doped crystals is lower than the output of the pure samples.

\section{Conclusion}
In the present paper results of measurements of absolute light yields on SrF$_2$, BaF$_2$ doped with various concentrations of Ce$^{3+}$ and Pr$^{3+}$ ions are reported. Pure strontium fluoride crystals demonstrate the highest light yield among fluorides. Photoelectron output of pure SrF$_2$ is higher than light output of CsI-Tl. Crystals doped with Ce$^{3+}$/Pr$^{3+}$ ions have lower light output than the pure ones.

\section*{Acknowledgment}
This work was partially supported by grant 11-02-00717 from Russian Foundation for Basic Research (RFBR). The study was also supported by The Ministry of education and science of Russian Federation. Authors are thankful to V. Kozlovskii for growing the crystals investigated in this work.
\bibliographystyle{IEEEtran}
\bibliography{IEEElib}

\end{document}